\documentclass[preprint,preprintnumbers,showpacs,amsmath,amssymb]{revtex4-1}

\usepackage{graphicx}% Include figure files
\usepackage{dcolumn}% Align table columns on decimal point
\usepackage{bm}% bold math

\begin{document}

%\preprint{PREPRINT L-1}

\title{Cross sections, multiplicity and moment distributions at the LHC}% Force line breaks with \\

% \title{Total cross section Hadron multiplicity at the LHC: the role of QCD effective charge}% Force line breaks with \\

\author{P.C. Beggio$^{1,2}$ and E.G.S. Luna$^{2}$ }
%\altaffiliation[Also at ]{Physics Department, XYZ University.}%Lines break automatically
%or can be forced with \\
\affiliation{
$^{1}$Laborat\'orio de Ci\^encias Matem\'aticas,Universidade Estadual do Norte
Fluminense Darcy Ribeiro, 28013-602, Campos dos Goytacazes, RJ, Brazil \\
$^{2}$Instituto de F\'{\i}sica, Universidade Federal do Rio Grande do Sul, CP 15051, 91501-970, Porto Alegre, RS, Brazil}


\begin{abstract}

The unitarity of the $S$-matrix requires that the absorptive part of the elastic scattering amplitude receives contributions
from both the inelastic and the elastic channels. We explore this unitarity condition in order to describe, in a connected way,
hadron-hadron observables like the total and elastic differential cross sections, the ratio of the real to imaginary part of
the forward scattering amplitude and the inclusive multiplicity distributions in full phase space,
over a large range of energies. We introduce non-perturbative QCD effects in the forward scattering amplitude
by using the infrared QCD effective charge dependent on the dynamical gluon mass. In our analysis we pay special
attention to the theoretical uncertainties in the predictions due to this mass scale variation.
We also present quantitative predictions for the $H_{q}$ moments at high energies. Our results reproduce the
moment oscillations observed in experimental data, and are consistent with the behavior predicted by QCD.

\end{abstract}

% \pacs{12.38.Lg, 13.85.Hd, 13.85.Lg}

%\pacs{}

\maketitle

\section{Introduction}

The study of multiple production of particles at high energies has been a subject of
intense theoretical and experimental interest. Its importance lies in the fact that the multiplicity distributions of
charged hadrons provide central information on the mechanism of production of the particles
\cite{dremin001,khoze001,khoze002,barroso}. Existing models
for particle production are
usually based on QCD since the production of hadrons can, at the microscopic level, be associated with the copious creation
of color partons through gluon radiation. However, this approach also contains a phenomenological component as the hadronization, the
transition of the quark-gluon system to hadrons, involves a soft scale, and therefore purely perturbative techniques do not apply
\cite{dremin001,khoze001,khoze002,Fiete,kittel003}.

The non-perturbative character of the QCD is also manifest in the elastic channel since at high energies the soft and the semihard
components of the scattering amplitude are closely related \cite{gribov001}. Thus, in considering the forward scattering amplitude for elastic
hadron-hadron collisions, it becomes important to distinguish between semihard gluons, which participate in hard parton-parton scattering, and
soft gluons, emitted in any given parton-parton QCD radiation process.

Fortunately, our task of describing hadron-hadron observables in both elastic and inelastic channels, bringing up information about the infrared
properties of QCD, can be properly addressed by considering the possibility that the non-perturbative dynamics of QCD generate an effective
gluon mass.
This dynamical gluon mass is intrinsically related to an infrared finite strong coupling constant, and its existence is strongly supported by
recent QCD lattice simulations \cite{lqcd} as well as by phenomenological results \cite{pheno001,luna008,luna009,luna010}.
More specifically, a global description of elastic
and inelastic hadronic observables can succeed in a consistent way by introducing a non-perturbative QCD effective charge in the calculation
of the gluon-gluon total cross section, which dominates at high energy and determines the asymptotic behavior of hadron-hadron cross sections,
and by exploring the unitarity condition of the $S$-matrix in impact parameter space, which relates the elastic scattering amplitudes to the
inelastic overlap function $G_{in}$. 

With this background in mind, the main purpose of this paper is to explore the non-perturbative dynamics of QCD in order to
describe in a connected way hadron-hadron observables in both elastic and inelastic channels,
assuming the eikonal representation and the unitarity condition of the scattering matrix, and compare the results with the $pp$ and
$\bar{p}p$ experimental data of total and elastic differential cross sections, the parameter $\rho$ and the inclusive multiplicity
distributions in full phase space.
% In our calculations the non-perturbative dynamics of QCD is introduced by using the infrared
% finite QCD effective charge dependent on the dynamical gluon mass.

The paper is organized as follows: in the next section we introduce a QCD-based eikonal model where the onset of the
dominance of semihard gluons in the interaction of high-energy hadrons is managed by the dynamical gluon mass.
Motivated by the recent TOTEM measurement of the $pp$ total cross section, $\sigma_{tot}^{pp}$, at LHC, we perform a detailed
analysis of $pp$ and $\bar{p}p$ forward scattering data and elastic differential cross sections using our eikonal model,
and obtain predictions for $\sigma_{tot}^{pp,\bar{p}p}$, $\rho^{pp,\bar{p}p}$ and $d\sigma^{\bar{p}p}/dt$ at Tevatron and
CERN-LHC energies. We evaluate the theoretical uncertainty associated with the
dynamical mass scale and obtain predictions with uncertainty bands for $\sigma_{tot}^{pp,\bar{p}p}$ and $\rho^{pp,\bar{p}p}$. 
In Sec. III we present the basic formalism, as well as the underlying physical picture, of multiplicity distributions
associated
with charged hadron production, and introduce the theoretical prescription for connecting the elastic and inelastic channels.
With the parameters of the eikonal fixed from the elastic fit, we calculate the multiplicity distribution $P_{n}$ in $pp$
and $\bar{p}p$ collisions over a large range of energies, and study the effect of the mass scale uncertainty on
high multiplicities.
In the sequence we calculate the $H_{q}$ moments of the
multiplicity distributions at high energies, where we observe that our results reproduce the oscillatory behavior predict
by QCD. In Sec. IV we draw our conclusions.

\section{Elastic channel: the Dynamical Gluon Mass model}

QCD-inspired models are at present one of the main theoretical approaches to explain the
observed increase of hadron-hadron total cross sections \cite{luna008,luna009,giulia001}. These models incorporate soft and
semihard processes in the treatment of high-energy hadron-hadron interactions using a formulation compatible with
analyticity and unitarity constraints. In the eikonal representation the total and inelastic cross sections are given by
\begin{eqnarray}
\sigma_{tot}(s)   =  4\pi   \int_{_{0}}^{^{\infty}}   \!\!  b\,   db\,
[1-e^{-\chi_{_{I}}(s,b)}\cos \chi_{_{R}}(s,b)],
\label{eq01}
\end{eqnarray}
\begin{eqnarray}
\sigma_{in}(s) = \sigma_{tot}(s) - \sigma_{el}(s) &=&  2\pi   \int_{_{0}}^{^{\infty}}   \!\!  b\,   db\,
G_{in}(s,b) \nonumber \\
 &=& 2\pi   \int_{_{0}}^{^{\infty}}   \!\!  b\,   db\,
[1-e^{-2\chi_{_{I}}(s,b)}],
\label{eq02}
\end{eqnarray}
respectively, where $s$ is the square of the total center-of-mass energy, $b$ is the impact parameter,
$G_{in}(s,b)$ is the inelastic overlap function and $\chi(s,b)=\chi_{_{R}}(s,b)+i\chi_{_{I}}(s,b)$
is the eikonal function. In this picture the probability that neither hadron is broken up in a collision
at impact parameter $b$ is therefore given by $P(s,b)=e^{-2\chi_{I}(s,b)}$. The eikonal function can
be written in terms of even and odd eikonal parts connected by crossing symmetry. Thus, in the case of
the proton-proton ($pp$) and antiproton-proton ($\bar{p}p$) scatterings, this combination reads
$\chi_{pp}^{\bar{p}p}(s,b) = \chi^{+} (s,b) \pm \chi^{-} (s,b)$. 

In the so called Dynamical Gluon Mass model \cite{luna008,luna009} 
(henceforth referred to as DGM model) 
the increase of the total cross sections is associated with semihard scatterings of partons in the hadrons, and
the high-energy dependence of the cross sections is driven mainly by gluon-gluon scattering processes. In this
QCD-inspired model
the even eikonal is written as the sum of gluon-gluon, quark-gluon, and quark-quark
contributions:
\begin{eqnarray}
\chi^{+}(s,b) &=&  \chi_{qq} (s,b) +\chi_{qg} (s,b)  + \chi_{gg} (s,b)
\nonumber  \\  &=&  i[\sigma_{qq}(s)  W(b;\mu_{qq})  +  \sigma_{qg}(s)
W(b;\mu_{qg}) + \sigma_{gg}(s) W(b;\mu_{gg})] ,
\label{eq22}
\end{eqnarray}
where $W(b;\mu)$ is the overlap density for the partons at impact parameter space $b$ and $\sigma_{ij}(s)$  are the
elementary subprocess cross sections of colliding quarks and gluons ($i,j=q,g$). The overlap density is
associated with the Fourier transform of a dipole form factor, $W(b;\mu) = \mu^{5} b^{3} \, K_{3}(\mu b)/96\pi$,
where $K_{3}(x)$ is  the modified Bessel function of  second kind. 
The contributions $\chi_{qq}  (s,b)$ and $\chi_{qg} (s,b)$ are parametrized with terms dictated by the Regge phenomenology:
\begin{eqnarray}
\chi_{qq}  (s,b) = i \Sigma \, A \, \frac{m_{g}}{\sqrt{s}} \, W(b;\mu_{qq}),
\end{eqnarray}
\begin{eqnarray}
\chi_{qg}  (s,b) = i \Sigma \left[ A' + B' \, \ln \left( \frac{s}{m_{g}^{2}} \right) \right] \, W(b;\mu_{qg}),
\end{eqnarray}
where $\mu_{qq}$, $\mu_{qg}$, $A$, $A'$ and $B'$ are fitting parameters. The $\Sigma$ factor is defined as
$\Sigma=9\pi\bar{\alpha}_{s}^{2}(0)/m_{g}^{2}$, where $\bar{\alpha}_{s}$ and $m_{g}$ are non-perturbative quantities which
will be defined in the ensuing text.
Since the role of the odd eikonal $\chi^{-}(s,b)$ is to account for the difference between $pp$ and $\bar{p}p$ channels at
low energies, it is simply written as
$\chi^{-} (s,b)  = C^{-}\, \Sigma \,  \frac{m_{g}}{\sqrt{s}} \, e^{i\pi/4}\, W(b;\mu^{-})$,
where $C^{-}$ and $\mu^{-}$ are constants to be fitted. The details concerning the analyticity properties of the model
amplitudes are given in Ref. \cite{luna008}.

The last term of the expression (\ref{eq22}), the gluon contribution $\chi_{gg}(s,b)= \sigma_{gg}(s)W(b; \mu_{gg})$,
deserves a more detailed comment: it gives the main contribution to the asymptotic behavior of hadron-hadron  total cross
sections; its energy dependence comes from the gluon-gluon cross section
\begin{eqnarray}
\sigma_{gg}(s) = C_{gg} \int_{4m_{g}^{2}/s}^{1} d\tau \,F_{gg}(\tau )\,
\hat{\sigma} (\hat{s}) ,
\label{eq28}
\end{eqnarray}
where $\tau = x_{1}x_{2} =\hat{s}/s$. Here $F_{gg}(\tau )\equiv [g\otimes g](\tau)=\int_{\tau}^{1} \frac{dx}{x}\, g(x)\,
g\left( \frac{\tau}{x}\right)$ is the convoluted structure function for pair gluon-gluon ($gg$), $\hat{\sigma} (\hat{s})$
is the total cross section for the subprocess $gg\rightarrow gg$, and $C_{gg}$ is a free parameter. Note that since $s \gg m_{g}$ the dominant contribution
to the integration comes from the small-$x$ region. In addition, the dynamical gluon mass scale is a natural regulator for the infrared divergences
associated with the gluon-gluon cross section. Since in the model the
small-$x$ semihard gluons play a central role, a phenomenological gluon distribution is introduced, namely
$g(x)=N_{g}\, (1-x)^{5}/x^{J}$, where $J=1+\epsilon$ and $N_{g}=\frac{1}{240}(6-\epsilon)(5-\epsilon)...(1-\epsilon)$.
Note that the gluon distribution reduces to $g(x)=0$ in the limit $x \to 1$, as expected by dimensional counting rules.
The total cross section $\hat{\sigma}(\hat{s})=\int_{\hat{t}_{min}}^{\hat{t}_{max}} (d\hat{\sigma}/d\hat{t}\, ) \, d\hat{t}$ is
given by \cite{luna008,luna009}
\begin{eqnarray}
\hat{\sigma}(\hat{s})              =              \frac{3\pi
  \bar{\alpha}_{s}^{2}}{\hat{s}}   \left[  \frac{12\hat{s}^{4}   -  55
  M_{g}^{2}  \hat{s}^{3} +  12  M_{g}^{4} \hat{s}^{2}  + 66  M_{g}^{6}
  \hat{s}   -   8 M_{g}^{8}}{4   M_{g}^{2}   \hat{s}   [\hat{s}   -
  M_{g}^{2}]^{2}}  -     3     \ln     \left(    \frac{\hat{s}     -
  3M_{g}^{2}}{M_{g}^{2}}\right) \right] ;
\label{h6}
\end{eqnarray}
in the above expression $\bar{\alpha}_{s}$ and $M_{g}$ are the QCD effective
charge and the dynamical gluon mass, respectively, obtained (through the use of the pinch
technique) as solutions of the Schwinger-Dyson equations for the gluon propagator and the triple gluon
vertex \cite{cornwall}. These non-perturbative effective quantities are given by
\begin{eqnarray} 
\bar{\alpha}_{s} = \bar{\alpha}_{s} (\hat{s})= \frac{4\pi}{\beta_0 \ln\left[
(\hat{s}^2 + 4M_g^2(\hat{s}^2) )/\Lambda^2 \right]}, 
\label{eq11}
\end{eqnarray}
\begin{eqnarray}
M^2_g = M^2_g(\hat{s}) =m_g^2 \left[\frac{ \ln
\left(\frac{\hat{s}^2+4{m_g}^2}{\Lambda ^2}\right) } {
\ln\left(\frac{4{m_g}^2}{\Lambda ^2}\right) }\right]^{- 12/11} ,
\label{mdyna} 
\end{eqnarray}
where $\Lambda$($\equiv\Lambda_{QCD}$)  is the QCD scale
parameter, $\beta_0 =  11- \frac{2}{3}n_f$  ($n_f$ is the
number of flavors), and $m_{g}$ is an infrared mass scale to be adjusted in order to provide reliable
results concerning calculations of strongly interacting processes. As mentioned in the earlier section, the existence of a
gluon mass scale is strongly supported by QCD lattice simulations \cite{lqcd}, and its value is typically found to be of
the order $m_{g}=500\pm200$ MeV \cite{pheno001,luna008,luna009}. It is
worth noting that in all these phenomenological studies the mass scale $m_{g}$ is used as an input parameter. However, it was
recently obtained, for the first time, a value for the mass scale $m_{g}$ from the analysis of the proton structure function
$F_{2}(x,Q^2)$ at small-$x$ \cite{luna010}. In that work the HERA data on $F_{2}$ is interpreted in terms of the generalized
double-asymptotic-scaling approximation with a QCD effective charge, and at leading order (LO) the value
$m_{g}=364\pm26$ MeV arises naturally from the analysis. This 90\% confidence level uncertainty result, obtained from a
global fit to $F_{2}(x,Q^2)$ data sets
% at low and moderate $Q^{2}$ values
through a $\chi^{2}$ fitting procedure, sets up the stage for making an assessment of the
mass scale uncertainty for the hadron-hadron observable to be calculated, as discussed in what follows.

\begin{table*}
\caption{Values of the DGM model parameters from the global fit to the scattering $pp$ and
$\bar{p}p$ data.}
% The dynamical gluon mass scale was set to $m_{g}=364$ MeV.}
\begin{ruledtabular}
\begin{tabular}{cc}
$C_{gg}$ & (1.62$\pm$0.37)$\times 10^{-3}$ \\
$\mu_{gg}$ [GeV]& 0.642$\pm$0.034 \\
$A$ & 9.04$\pm$4.94 \\
$\mu_{qq}$ [GeV] & 1.299$\pm$0.797 \\
$A^{\prime}$ & (4.68$\pm$1.89)$\times 10^{-1}$ \\
$B^{\prime}$ & (4.53$\pm$1.94)$\times 10^{-2}$ \\
$\mu_{qg}$ [GeV] & 0.825 $\pm$0.015 \\
$C^{-}$ & 3.12$\pm$0.33 \\
$\mu^{-}$ [GeV]& 0.799$\pm$0.298 \\
\hline
$\chi^{2}/DOF$  &  0.98 \\
\end{tabular}
\end{ruledtabular}
\end{table*}

First, in order to determine the DGM model parameters, we set the value of the gluon scale mass to $m_{g} = 364$ MeV, and
fix $n_f = 4$ and $\Lambda = 284$ MeV. These choices are not only consistent to LO procedures, but are also the same ones
adopted in Ref. \cite{luna010}. Then we carry out a global fit to all high-energy forward $pp$ and $\bar{p}p$ scattering data above
$\sqrt{s} = 10$ GeV, namely the total cross section $\sigma_{tot}^{pp,\bar{p}p}$, the ratio of the real to
imaginary part of the forward scattering amplitude $\rho^{pp,\bar{p}p}$, the elastic differential scattering cross sections
$d\sigma^{\bar{p}p}/dt$ at $\sqrt{s} = 546$ GeV and $\sqrt{s} = 1.8$ TeV as well as the recent TOTEM datum on
$\sigma_{tot}^{pp}$ at $\sqrt{s} = 7$ TeV
\cite{totem}. The values of the fitted parameters are given in Table 1. The $\chi^{2}/DOF$ for the global fit was 0.98 for
320 degrees of freedom. This result is obtained by fixing the quantity $J = 1 +\epsilon$, which determines the asymptotic
behavior of $\sigma_{tot}^{pp}$, at $J=1.21$; it appears to be the optimal value to fit the TOTEM datum on $\sigma_{tot}^{pp}$ at
$\sqrt{s} = 7$ TeV. Interestingly enough, the value $J=1.21$ is the same value obtained from a recent triple-Regge analysis which
explicitly accounts for absorptive corrections \cite{lunakmr001}.
The results of the fits to $\sigma_{tot}^{pp,\bar{p}p}$, $\rho^{pp,\bar{p}p}$ and $d\sigma^{\bar{p}p}/dt$
are displayed in Figs. 1a, 1b and 1c, respectively, together with the experimental data. The hatched areas in Figs. 1a and
1b correspond to the uncertainties in the predictions due to the gluon mass error, namely $\delta m_{g}=26$ MeV. 
In Figure 1d we show the theoretical predictions for the $pp$ differential scattering cross sections at $\sqrt{s} = 7$ and 14 TeV; the
comparison of the prediction at $\sqrt{s} = 7$ TeV with the published experimental data \cite{totem} shows good agreement.

\section{Inelastic channel: the string approach}

The number $n$ of charged hadrons in the final state of a collision is a basic observable in high-energy processes. Although
$n$ is one of the simplest observables in hadron-hadron collisions, being a global measure characterizing final states, it 
provides important constraints on the mechanism of production of the particles. Accordingly, the
charged particle multiplicity distribution $P_{n}$, which give us the probability for producing $n$ charged hadrons in a given event,
is the most general characteristic of multi-particle production processes. The multiplicity distribution $P_{n}$ follows a
Poisson distribution in the case of independent production of the final-state particles, but deviations from the Poisson shape
contains information about multi-particle correlations \cite{dremin001,Fiete}.
Higher-order moments of $P_{n}$, which measure the event-to-event multiplicity fluctuations, constitute a powerful tool
for studying these correlations.
Measurements of the $s$ dependence of $P_{n}$ and its moments has been used to improve or reject models of particle production.

The multiplicity distributions are usually represented in terms of the Koba-Nielsen-Olesen (KNO) function
$\Phi(z)\equiv \langle n \rangle P_{n}$, where $\langle n \rangle$ is the mean of the multiplicity
distribution and $z=n/\langle n \rangle$. It is experimentally observed that for hadron-hadron collisions the KNO function
$\Phi(z)$ is approximately independent of $s$ (KNO scaling) over the ISR energy range and shows
dramatic scaling violation from ISR to FNAL and LHC energies. In these collisions the production of minijets from semihard
partonic processes becomes increasingly important at high energies, and in many theoretical approaches the minijets are
the responsible not only for the rapid growth of hadron-hadron cross sections but also for the violation of the KNO scaling.
This renders the DGM model particularly adept at calculating $P_{n}$ and its moments, since in the model
the increase of the $pp$ and $\bar{p}p$ total cross sections as well as the violation of KNO scaling is driven especially by
the semihard component of the eikonal model \cite{luna008,luna009,Beggio001,BeggioNPA}.

A reliable approach for the calculation of the multiplicity distribution, using the DGM model, can be established by
exploring the unitarity condition of the $S$-matrix in the following way:
% first we observe that
the elastic scattering amplitude, in the impact parameter $b$ space, may be written
\begin{eqnarray}
2 \textnormal{Re} \Gamma (s,b) = |\Gamma (s,b)|^{2} + G_{in}(s,b) ,
\label{unitcond}
\end{eqnarray}
where $\Gamma (s,b)$ is the profile function and $G_{in}(s,b)$ the inelastic overlap function. From the optical
theorem we obtain $\sigma_{tot}(s)=2\int d^{2}b\, \textnormal{Re} \Gamma (s,b)$, and since the function $G_{in}(s,b)$ represents the
probability of absorption associated to each $b$ value, we get the total inelastic cross section
$\sigma_{in}(s) = \int d^{2}b \, G_{in}(s,b)$. Thus the unitarity condition (\ref{unitcond}) is equivalent to 
$\sigma_{tot}(s)=\sigma_{el}(s)+\sigma_{in}(s)$, where $\sigma_{el}(s) = \int d^{2}b \, |\Gamma (s,b)|^{2}$. The multiplicity
distribution $P_{n}(s)$, the probability of producing $n$ charged particles in an inelastic collision at the energy $s$, is given by
\begin{eqnarray}
P_{n}(s) = \frac{\sigma_{n}(s)}{\sigma_{in}(s)} ,
\end{eqnarray}
where $\sigma_{n}(s)$ is the $n$-particle topological cross section, with $\sum_{n} \sigma_{n}(s) = \sigma_{in}(s)$. The
connection between the multiplicity distribution and the DGM eikonal becomes complete using a phenomenological procedure
referred to as {\it geometrical} or {\it string approach} \cite{Beggio001,BeggioNPA,BeggioMV,BeggioHama,BeggioBPJ}, where the
hadronic multiplicity $P_{n}(s)$ can be constructed in terms of elementary quantities related to microscopic processes.
More specifically, the topological cross section is written as
\begin{eqnarray}
\sigma_{n}(s) \equiv \int d^{2}b \, \sigma_{n}(s,b) = \int d^{2}b \, G_{in}(s,b) \left[ \frac{\sigma_{n}(s,b)}{\sigma_{in}(s,b)} \right]  ,
\end{eqnarray}
since $\sigma_{in}(s)\equiv \int d^{2}b \, \sigma_{in}(s,b) = \int d^{2}b \, G_{in}(s,b)$. In this way the multiplicity
distribution $P_{n}(s)$ is given by
\begin{eqnarray}
P_{n}(s) = \frac{1}{\sigma_{in}(s)} \int d^{2}b \, \frac{G_{in}(s,b)}{\langle n(s,b) \rangle }
\left[\langle n(s,b) \rangle p_{n}(s,b) \right]  ,
\label{distrib01}
\end{eqnarray}
where $\langle n(s,b) \rangle$ is the average number of particles produced at $b$ and $s$, and
$p_{n}(s,b)\equiv \sigma_{n}(s,b)/\sigma_{in}(s,b)$.
% due to interactions among elementary constituents of hadrons involved in the collision.
Let
$\Phi(s,z)=\langle n(s) \rangle P_{n}(s)$ be the overall multiplicity distribution, where
$z=n(s)/\langle n(s) \rangle $ is the corresponding KNO variable. A multiplicity distribution $\phi(s,\b{z})$, associated
with elementary
processes occurring at $b$ and $s$, can be written in the form $\phi(s,\b{z}) = \langle n(s,b) \rangle p_{n}(s,b)$, where
$\b{z} = n(s)/\langle n(s,b) \rangle$. Note that both distributions obey the usual normalizations \cite{barshay01,lam01}
\begin{eqnarray}
\int_{0}^{\infty} dz\, \Phi(z) = \int_{0}^{\infty} d\b{z}\, \phi(\b{z}) = 2
\label{normaliz001}
\end{eqnarray}
and
\begin{eqnarray}
\int_{0}^{\infty} dz\, z\, \Phi(z) = \int_{0}^{\infty} d\b{z}\, \b{z}\, \phi(\b{z}) = 2 .
\label{normaliz002}
\end{eqnarray}
In this approach the average number of particles $\langle n(s,b) \rangle$ factorizes as
$\langle n(s,b) \rangle = \langle n(s) \rangle f(s,b)$, and therefore the expression (\ref{distrib01}) can be rewritten in the KNO form
\begin{eqnarray}
\Phi (s,z) = \langle n(s) \rangle P_{n}(s) = \frac{1}{\int d^{2}b \, G_{in}(s,b)} \int d^{2}b \, \frac{G_{in}(s,b)}{f(s,b)}\,
\phi \left( \frac{z}{f(s,b)} \right) ,
\label{distrib02}
\end{eqnarray}
where $f(s,b)$ is a {\it multiplicity function} to be defined. Thus the connection between $\Phi (s,z)$ and the DGM eikonal can finally be
established since, as shown in the expression (\ref{eq02}),
the inelastic overlap function is simply $G_{in}(s,b)=1-e^{-2\chi_{_{I}}(s,b)}$. Therefore
\begin{eqnarray}
P_{n}(s) = \frac{1}{\langle n(s) \rangle\int d^{2}b \, \left[ 1-e^{-2\chi_{_{I}}(s,b)} \right] }
\int d^{2}b \, \frac{\left[ 1-e^{-2\chi_{_{I}}(s,b)} \right]}{f(s,b)}\,
\phi^{(1)} \left( \frac{z}{f(s,b)} \right) ,
\label{distrib03}
\end{eqnarray}
where we have introduced an index labeling the function $\phi$ in order to stress the underlying assumption of the string
approach: the hadronic multiplicity $P_{n}(s)$ is obtained by summing contributions from parton-parton interactions
occurring at fixed $b$ and $s$, with the subsequent creation, per collision, of an object like a string. These strings give
rise to parton cascades followed by the non-perturbative hadronization phase transition, where the partons become the
hadrons which are experimentally observed. In our case each string has equal probability of turning into a pair of charged
hadrons.

It is well known that over a large range of energies the experimental data on multiplicity distributions are well fitted by the Negative Binomial
Distribution (NBD). Hence, in this work we take the KNO form of the NBD, or Gamma distribution,
\begin{equation}
\phi^{(1)}(z)=2 \,\frac{K^K}{\Gamma(K)} \, z^{K-1} \, e^{-Kz} ,
\label{psi1}
\end{equation}
as the parametrization of the elementary multiplicity distribution $\phi^{(1)}$. This choice allows for a
natural interpretation of the parameter $K$ as a dimensionless minimum length for a multitude of string-like objects, as discussed by Barshay
many years ago \cite{Barshay001}. Moreover, the Gamma distribution arises as the dominant part of the solution of the equation for three-gluon
branching in the very-large-$n$ limit \cite{durand001}; this branching equation, which takes into account only gluon 
bremsstrahlung process, gives the main contribution at high energies since semihard gluons dominate the parton-parton cross sections.
Curiously enough, from the color glass condensate approach the NBD appears to be the underlying distribution of gluon multiplicities arising
from electric and magnetic flux tubes in the Glasma \cite{gelis001}.

In addiction, it has been assumed that the energy effectively employed for the production of particles in a
given interaction is proportional to the eikonal \cite{Beggio001,BeggioNPA,BeggioMV,BeggioHama,BeggioBPJ,barshay01,Barshay001}. It means
that the eikonal
may be interpreted as a measure of the effective overlap of two colliding matter distributions on the impact parameter plane.
As a result, since the high-energy eikonal is
largely imaginary, we adopt the following form for the multiplicity function $f(s,b)$:
\begin{equation}
f(s,b)=\xi(s)\,[\chi_{I}(s,b)]^{2\zeta} ,
\label{xi001}
\end{equation}
where the relation 
\begin{equation}
 \xi(s)=\frac
  {\int d^2 b\,[1-e^{-2\,\chi_{I}(s,b)}]}
  {\int d^2 b\,[1-e^{-2\,\chi_{I}(s,b)}]\,[\chi_{I}(s,b)]^{2\zeta}}
 \label{xi002}
\end{equation}
is determined by the normalization condition (\ref{normaliz002}). Since
the eikonal $\chi_{I}(s,b)$ is completely determined from the elastic fit, we see from Eqs. (\ref{psi1}) and (\ref{xi001}) that
the only free parameters in Eq. (\ref{distrib03}) are $K$ and $\zeta$. Further, for our purposes it is
sufficient to fix the value of $K$ and assume $\zeta$ as the single fitting parameter; in Refs. \cite{BeggioNPA} and \cite{BeggioMV} it
is shown that the choice $K=10.775$ gives good descriptions of the charged multiplicity distributions for $e^{+}e^{-}$
annihilation data in a large interval of $\sqrt{s_{e^{+}e^{-}}}$. Moreover, our assumption that each string has equal
probability of turning into hadrons suggests a common minimum length $K$ for different $pp$ and $\bar{p}p$ centre-of-mass
energies. More important, this procedure in fact does provide an excellent description of $P_{n}$ data at high multiplicities,
avoiding the introduction of more free parameters, as it is shown in the sequence.

In order to determine the parameter $\zeta$, we set the value of the dimensionless length to $K=10.775$, and carry out fits to
full phase space $P_{n}$ data over a large range of energies, namely at
$\sqrt{s}=30.4, 44.5, 52.6, 62.2, 300, 546, 1000$ and $1800$ GeV \cite{ABC,Alexopoulos}. For energies $\sqrt{s}\geq 300$ GeV
we have used data from the E735 Collaboration since it is statistically more reliable in the high multiplicity region
\cite{Alexopoulos}. It is worth mentioning that it is the region clearly sensitive to the gluon distribution at small-$x$, and
therefore to the underlying mechanisms of the DGM model.
In all these fits we use a $\chi^{2}$ fitting procedure with an interval $\chi^{2}-\chi^{2}_{min}$ corresponding to the projection of
the $\chi^{2}$ hypersurface containing 90\% of probability. The parameter values obtained in each fit,
as well as the associated $\xi(s)$ and $\langle n(s)\rangle$ values, are given in Table II.

%% \begin{table}[htb!]
\begin{table*}
\caption{Values of the $\zeta$ parameter resulting from fits to the $P_{n}$ data. The values of $\xi(s)$ were obtained from
Eq. (\ref{xi002}). At ISR
energies the values of $\langle n(s) \rangle$ were inputed from Ref. \cite{ABC}, while in the other energies
have been obtained from empirical $P_{n}$ data by using $\langle n \rangle=\sum n\,P_{n}$. At LHC energies the
values were estimated from Eq. (\ref{troshin001}).}
\begin{ruledtabular}
\begin{tabular}{ccccc}
% \hline
$\sqrt{s}$ [GeV] & $\zeta$ & $\xi(s)$ & $\langle n(s) \rangle $ & $\chi^{2}/DOF$ \\
\hline
30.4 & 0.239 $\pm$ 0.011 & $1.642$ & $ 9.43$ & $0.588$ \\
44.5 & 0.240 $\pm$ 0.011 & $1.643$ & $10.86$ & $0.306$ \\
52.6 & 0.239 $\pm$ 0.009 & $1.639$ & $11.55$ & $0.765$ \\
62.2 & 0.231 $\pm$ 0.008 & $1.613$ & $12.25$ & $1.717$ \\
300  & 0.263 $\pm$ 0.003 & $1.589$ & $24.47$ & $0.608$ \\
546  & 0.305 $\pm$ 0.004 & $1.599$ & $29.53$ & $0.300$ \\
1000 & 0.288 $\pm$ 0.005 & $1.508$ & $38.46$ & $1.469$ \\
1800 & 0.315 $\pm$ 0.002 & $1.468$ & $44.82$ & $0.782$ \\
7000 & 0.352             & $1.308$ & $81.79$ & $  -   $ \\
14000 & 0.372            & $1.209$ & $108  $ & $  -   $ \\
% \hline
\end{tabular}
\end{ruledtabular}
\end{table*}

The theoretical prediction to $P_{n}$ at LHC energies can be computed by fitting the energy behavior of $\zeta$
through an appropriate function $\zeta(s)$. As can be seen from the Figure 2, the parameter increases rapidly with $s$, being its energy dependence
described in a consistent way through the use of the function
\begin{eqnarray}
\zeta(s)=a+b \ln^{c}(s),
\end{eqnarray}
where $a$=0.189, $b$=0.00197 and $c$=1.536; these values were obtained by a $\chi^{2}$ analysis. 
From the curve depicted in Fig. 2 it is possible to extract the values of $\zeta$ at 7 and 14 TeV shown in the Table II. In the same way, the
values of the hadronic average multiplicity at LHC energies can be estimated using the equation introduced by Troshin and Tyurin \cite{Troshin}
\begin{equation}
\langle n(s)\rangle = 2.328\,s^{0.201},
\label{troshin001}
\end{equation}
which is in very good agreement with experimental data.  

The results of the fits to $P_{n}$ data are displayed in Figs. 3 and 4, together with the experimental data. In Figure 4 the hatched areas
correspond to the uncertainties in the predictions due to the gluon mass error. Notice that, since at ISR energies gluon-gluon processes
are not dominant, the widths of uncertainties bands in Figure 3 are very narrow. In Figure 5 we show the theoretical predictions for the
full-phase-space multiplicity distribution at $\sqrt{s}=7$ and 14 TeV, together with data, from CMS, on fully corrected charged hadron
multiplicity spectrum for $|\eta|<0.5$, 1.0, 1.5, 2.0 and 2.4 at $\sqrt{s}=7$ TeV \cite{CMS}.

Information about multi-particle correlations can be extracted directly from the shape of the multiplicity distribution $P_{n}$. As first
indicated in the Section III, independent production of the particles in the final state leads to a Poissonian multiplicity
distribution. However, deviations from this Poisson shape reveal correlations among the produced particles and unveil many details
about the underlying dynamics.

The shape of a distribution can be analyzed by studying its moments. In the case of the multiplicity
distribution $P_{n}$, its factorial moments $F_{q}$ as well as its cumulant factorial moments $K_{q}$ are powerful tools to investigate
multiplicity fluctuations \cite{dremin001,kittel001,GUNPB,L3001,sarkisyan001}. The normalized factorial moment of rank $q$ is defined by
\begin{equation}
F_{q}=\frac{1}{\langle n\rangle^{q}} \sum_{n=q}^{\infty}
n(n-1)...(n-q+1)\,P_{n}\,,
\label{fatorial}
\end{equation}
where $\langle n\rangle =\sum_{n} nP_{n}$ is the average multiplicity and $q$ is the rank of the moment. Thus $F_{q}$,
that reflects correlations among up to $q$ produced particles, is sensitive to the tail of the distribution, i.e. to large $n$ values.
For independent production of particles the moment $F_{q}$ is equal to unity for all rank $q$. On the other hand, if the
produced particles are correlated (anti-correlated), the distribution is broader (narrower) than the Poisson one, and,
as a consequence, $F_{q}$ is greater (less) than unity.
The normalized factorial cumulants $K_{q}$, obtained from the recursive relation
\begin{equation}
K_{q} = F_{q} - \sum_{m=1}^{q-1} \frac{(q-1)!}{m!(q-m-1)!}\,K_{q-m}\,F_{m} 
 \label{cumulant}
\end{equation}
with $F_{0}=F_{1}=K_{1}=1$, measure genuine $q$-particle correlations, more specifically correlations of rank $q$ which are not a
consequence of lower-order correlation functions. It is observed that the factorial and cumulant moments increase rapidly with $q$ and,
therefore, the ratio \cite{DreminNechitailo001,Dremin002}
\begin{equation}
H_{q}=\frac{K_{q}}{F_{q}}\,
\label{hq}
\end{equation}
turns out to be a convenient quantity to work with since it have a flatter behavior with $q$ over a large range.
Moreover, higher-order QCD calculations predict that the cumulant and $H_{q}$ moments oscillate as a function of $q$
\cite{DreminNechitailo001,Dremin002,Dremin003}. These predictions have been fully confirmed by high-energy scattering experiments
in $e^{+}e^{-}$, $pp$, $\bar{p}p$, hadron-nucleus and nucleus-nucleus collisions \cite{DiasdeDeus}. Namely, QCD results are concerned with
partons, not with hadrons. In fact, the $H_{q}$ moments of {\it parton} multiplicity distributions are predicted to oscillate. However,
$H_{q}$ moments of experimental data on {\it hadron} multiplicity distributions exhibit the same oscillatory behavior. This experimental
observation strongly suggests the validity of the Local Parton Hadron Duality (LPHD) hypothesis \cite{LPHD001,LPHD002}, which asserts
that parton multiplicity distributions, calculated using perturbative QCD, must be related to the hadronic distributions, 
implying that the hadronization phase transition does not affect in an essential way the observed final-state hadron distributions.

We have analyzed the $H_{q}$ moments over a large range of the rank $q$. First, the theoretical values of $P_{n}$ are obtained
from the Eqs. (\ref{distrib03})-(\ref{xi002}). Then the $F_{q}$, $K_{q}$ and $H_{q}$ moments are determined from both the
theoretical and the experimental values of $P_{n}$ according to Eqs. (\ref{fatorial})-(\ref{hq}).
The results are displayed in Figs. 6 and 7. Our prediction for the $H_{q}$ moment at the LHC energy $\sqrt{s}=14$ TeV is presented in Fig. 8.
All these results are in good agreement with the experimental points up to $q$=16, i.e., are reliable even on the tail of the distribution.

\section{Concluding Remarks}

In this paper we have explored the unitarity condition of the $S$-matrix in impact parameter space to
describe, in a connected way, hadron-hadron observables in both elastic and inelastic channels.
In the elastic channel we have performed a global fit to the $pp$ and $\bar{p}p$ forward quantities $\sigma_{tot}$ and $\rho$, including
in the dataset the $\bar{p}p$ differential cross section at $\sqrt{s}=546$ GeV and $\sqrt{s}=1.8$ TeV as well as the new TOTEM datum on
$\sigma_{tot}^{pp}$ at $\sqrt{s}=7$ TeV. The analysis was carried out using the DGM eikonal model, where non-perturbative effects are
naturally included via a QCD effective charge dependent on the dynamical gluon mass. With the dynamical gluon mass set at $m_{g} = 364$ MeV, the
DGM model allows us to describe successfully all experimental data. Namely, the $\chi^{2}/DOF$ for the global fit was 0.98 for 320 degrees of
freedom. This good statistical result shows that the DGM model is well suited for detailed predictions of the
forward quantities to be measured at higher energies. In particular, for the total cross sections to be
measured at CERN-LHC energies, the model predicts the values $\sigma_{tot}^{pp}=98.6^{+6.8}_{-5.2}$ mb, $101.1^{+7.0}_{-5.4}$ mb and
$112.4^{+8.1}_{-6.4}$ mb, at $\sqrt{s}=7$, 8 and 14 TeV, respectively. The uncertainty in these total cross sections have been estimated by
varying the gluon mass scale
within error while keeping all other model parameters constant. Note that this procedure does not determines the {\it formal} uncertainty 
in the $\sigma_{tot}$ prediction, since the variance-covariance matrix method, necessary for proving this quantity, was not employed. However,
at high energies the total cross sections are dominated by gluon-gluon interactions represented by the eikonal term $\chi_{gg}(s,b)$, containing
only 2 free parameters ($C_{gg}$ and $\mu_{gg}$). It is seen from our $\chi^{2}$ analysis that the correlation coefficients of these parameters
are small. Moreover, the values of the total cross sections are actually more sensitive to the gluon mass scale than to variations of others
parameters of the model. Hence, despite the simplicity of the procedure, it clearly provides a good estimate of the full systematic uncertainty
in $\sigma_{tot}$.

In order to compute multiplicity distributions in $pp$ and $\bar{p}p$ collisions we have adopted the string approach, which enables
linkage between the elastic and inelastic channels. This phenomenological approach asserts that $P_{n}$ in full phase space can be constructed by
summing contributions from parton-parton collisions, occurring at $b$ and $s$. These collisions give rise to the formation of strings and its
subsequent fragmentation into hadrons. More specifically, the semihard partons produced at the interaction point fly away from each other yielding
a color string which breaks up producing the observed hadrons, a process well described by the conventional Lund model \cite{lund001}.
Thus in this picture the multiplicity distribution $P_{n}$ can be constructed in terms of the eikonal $\chi_{I}(s,b)$, which contains the gluon
semihard contribution $\chi_{gg}(s,b)$, and of the parameters $K$ and $\zeta$, characterizing the dimensionless length of string-like objects and
the {\it strength} of the KNO scaling violation, respectively. In fact, the parameter $\zeta$ is approximately constant at ISR energies, increasing
rapidly with $s$ from ISR to LHC energies: this behavior is consistent with KNO scaling hypothesis at $\sqrt{s}\lesssim 100$ GeV and it's
violation at higher energies. We call attention in particular to the fact that the multiplicity
distribution data are fitted adequately using only one free parameter, since the $K$ as well as the elastic parameters are all fixed.

We point out that the bumpy structure in $P_{n}(s)$ data at higher energies, which appears in
the region of low multiplicities, can be described by our model by means of a moment analysis which separates the semihard
and soft components of the multiplicity distributions. In this analysis the bumpy structure is reproduced after
an appropriate convolution of the soft and the semihard components of the eikonal. However, at least one more free
parameter is needed. Since our approach aims to achieve a description of the data relying on the
smallest number of parameters possible, the description of the bumpy structure is out of the scope of this work.

Concerning the moments and cumulants, we see that our approach describes very well the energy dependence of the
$F$-moments, and reproduce the $H_{q}$ versus $q$ oscillations observed in the experimental data. More importantly, we observe that
the semihard portion of the eikonal is in fact the responsible for the $H_{q}$ oscillations since it gives the main contribution to high
multiplicities, as qualitatively predicted by QCD.

It is worth noting that the unitarity condition of the $S$-matrix can be also associated with low-$x$
saturation effects of $xG(x,Q^{2},|\vec{b}|)$, the $b$-dependent integrated gluon distribution \cite{shoshi001}. These saturation effects
appears in a functional integral approach to high-energy scattering in the eikonal representation, and the $S$-matrix unitarity is preserved
as the result of a matrix cumulant expansion and the Gaussian approximation of the functional integrals. In this non-perturbative approach an
{\it effective} gluon mass is introduced, indicating that the underlying physics is similar to that of the DGM model. It corroborates the idea that
the non-perturbative dynamics of QCD generate an effective gluon mass at very low $Q^{2}$ region.

\begin{acknowledgments}
We thank V.A. Khoze, M.J. Menon, A.A. Natale, F.S. Navarra and E.K.G. Sarkisyan
for helpful discussions. This research was partially supported by the Funda\c{c}\~{a}o de Amparo \`{a} Pesquisa do Estado do Rio Grande do Sul
(FAPERGS) and by Coordena\c{c}\~ao de Aperfei\c{c}oamento de Pessoal de N\'{\i}vel Superior (CAPES).
\end{acknowledgments}

\newpage

\begin{figure}
\label{difdad}
\vspace{2.0cm}
\begin{center}
%\vspace{-0.6cm}
\includegraphics[height=.60\textheight]{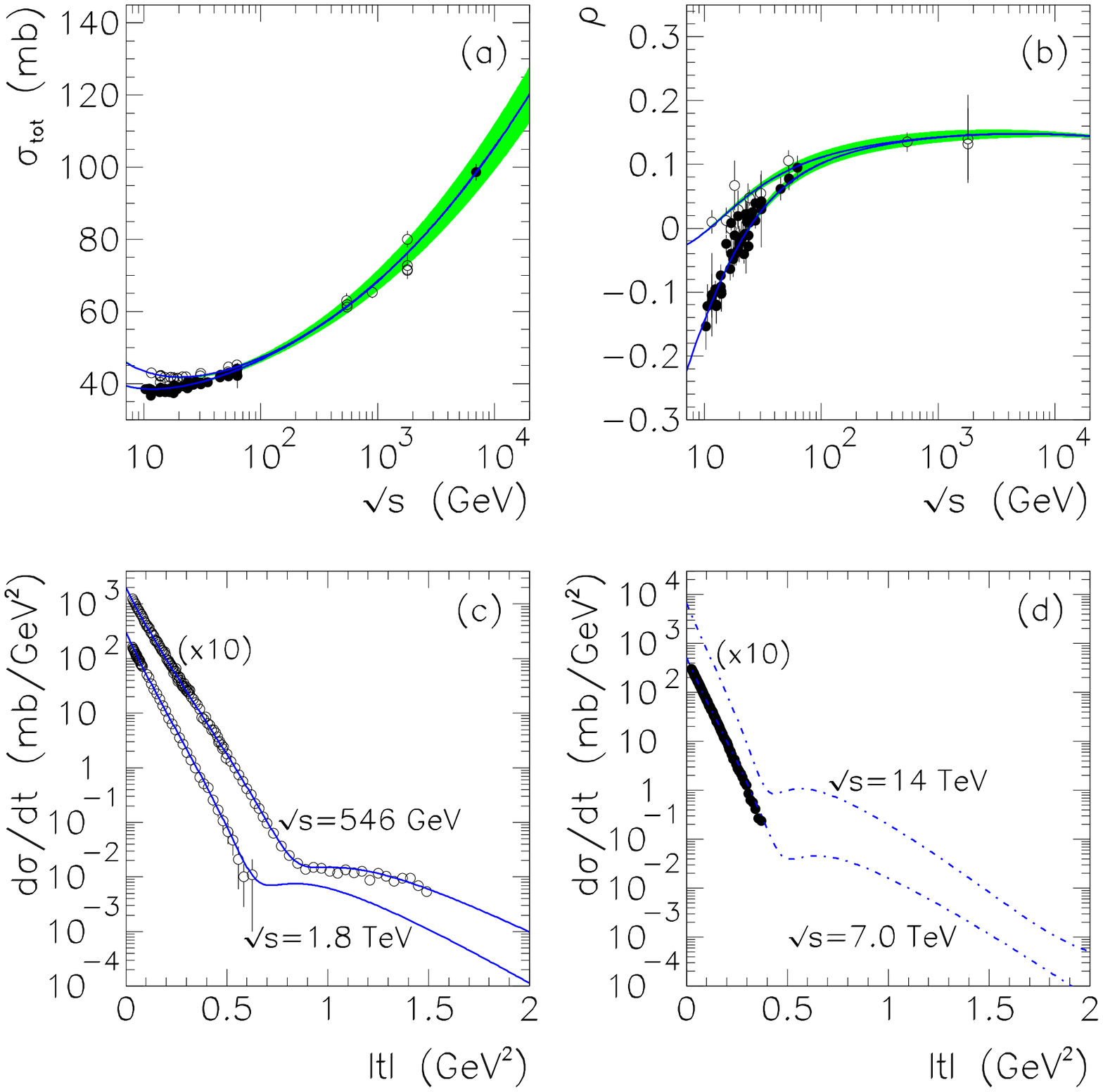}
\caption{(a) Total cross section, (b) ratio of the real to imaginary part of the forward scattering amplitude, and (c)
elastic differential scattering cross sections for $pp$ ($\bullet$) and $\bar{p}p$ ($\circ$). In (d) we show the theoretical
predictions for $d\sigma^{pp}/dt$ at $\sqrt{s} = 7$ and 14 TeV. The hatched areas in (a) and (b) correspond to the uncertainties in the
predictions due to the gluon mass error.}
\end{center}
\end{figure}

\begin{figure}
\label{difdad}
\vspace{2.0cm}
\begin{center}
%\vspace{-0.6cm}
\includegraphics[height=.60\textheight]{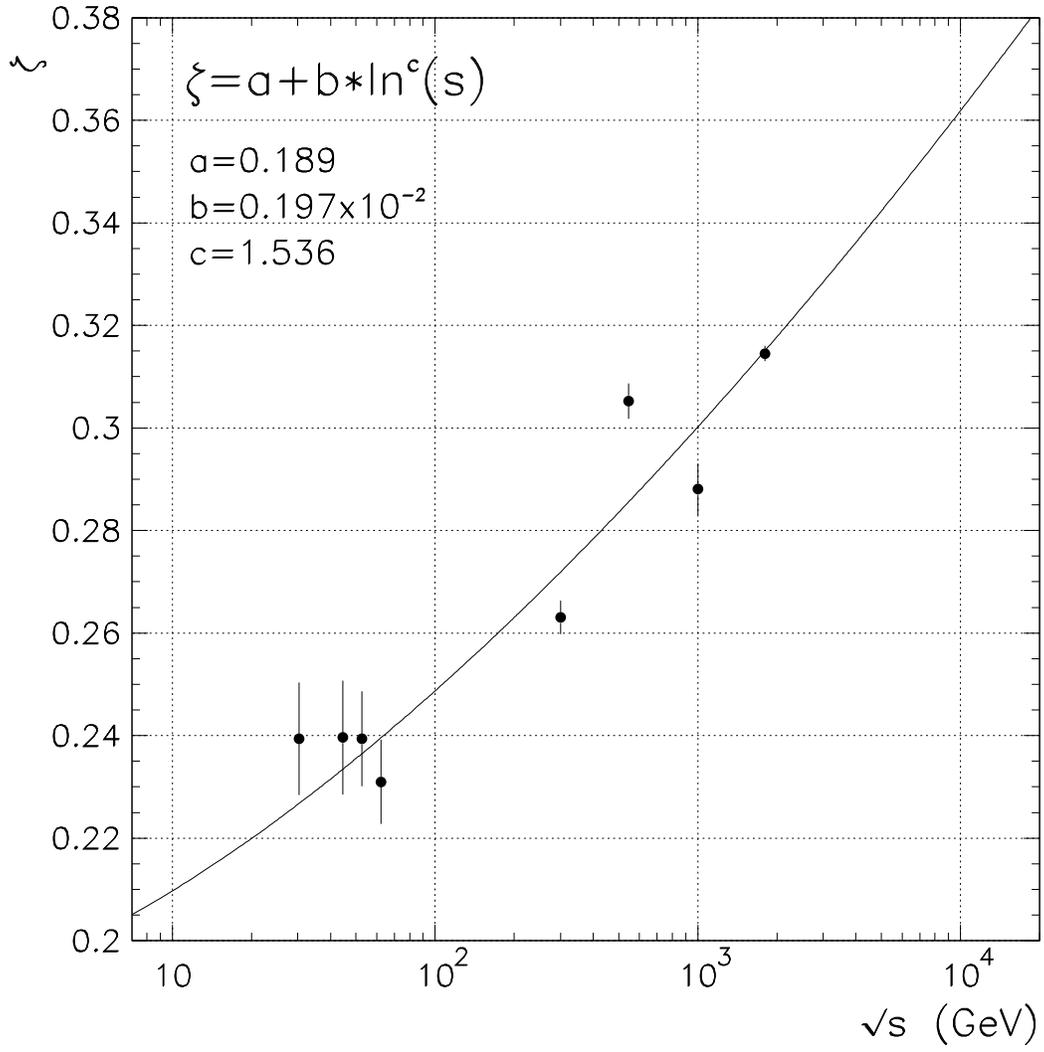}
\caption{The energy dependence of the parameter $\zeta$. }
\end{center}
\end{figure}

\begin{figure}
\label{difdad}
\vspace{2.0cm}
\begin{center}
%\vspace{-0.6cm}
\includegraphics[height=.60\textheight]{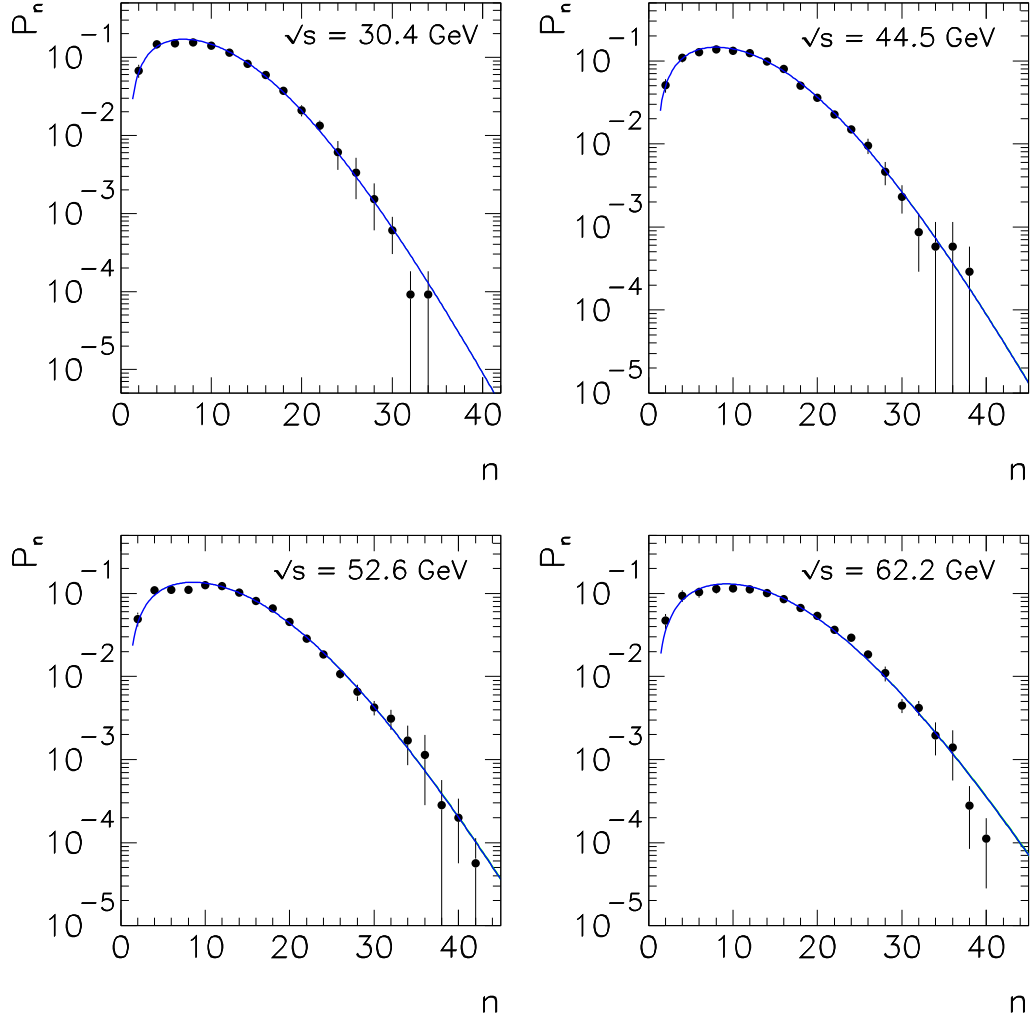}
\caption{Multiplicity distributions for inelastic $pp$ data at $\sqrt{s}=30.4, 44.5, 52.6$ and $62.2$ GeV compared with
theoretical expectations. }
\end{center}
\end{figure}

\begin{figure}
\label{difdad}
\vspace{2.0cm}
\begin{center}
%\vspace{-0.6cm}
\includegraphics[height=.60\textheight]{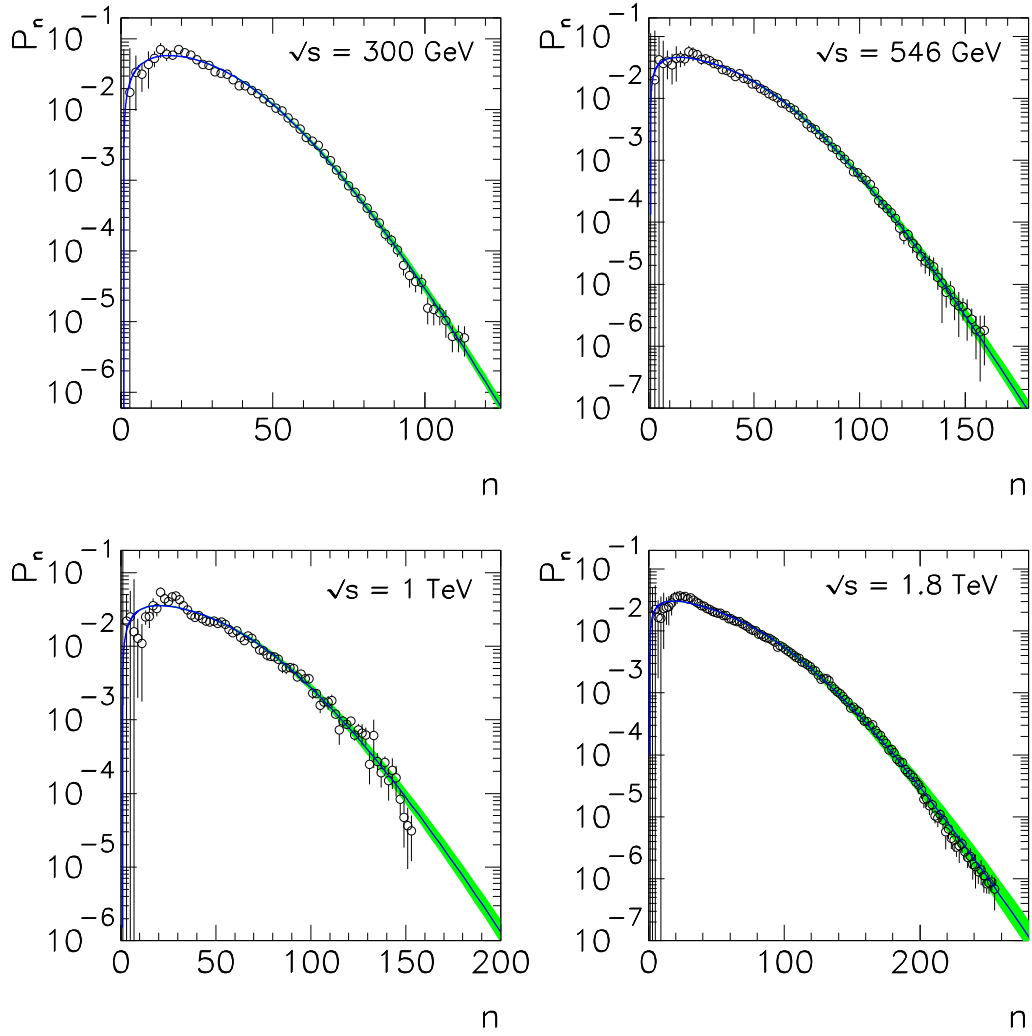}
\caption{Multiplicity distributions for inelastic $\bar{p}p$ data at $\sqrt{s}=300, 546, 1000$ and $1800$ GeV compared with
theoretical expectations. The hatched areas correspond to the uncertainties in the predictions due to the gluon mass error.}
\end{center}
\end{figure}

\begin{figure}
\label{difdad009}
\vspace{2.0cm}
\begin{center}
%\vspace{-0.6cm}
\includegraphics[height=.60\textheight]{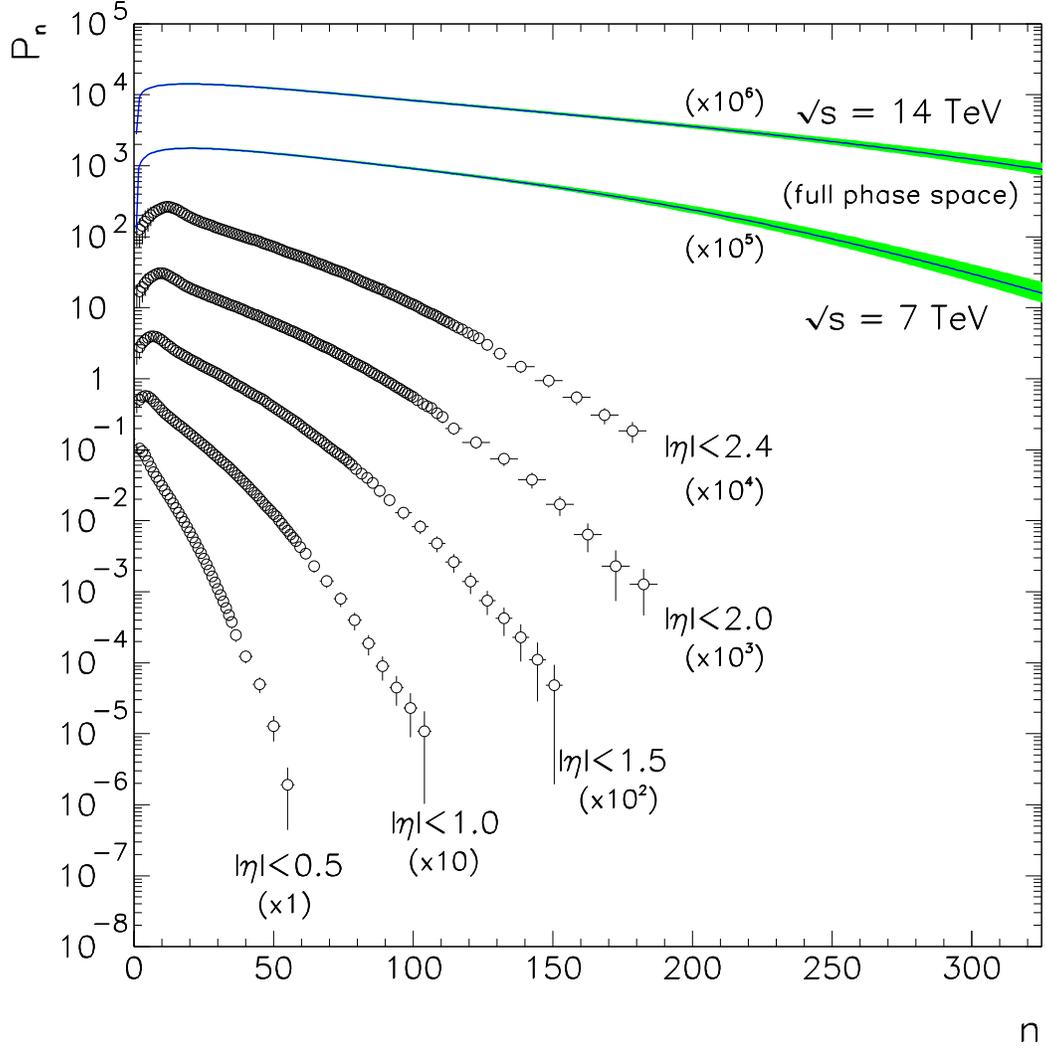}
\caption{Theoretical predictions for the full-phase-space multiplicity distribution at $\sqrt{s}=7$ and 14 TeV, together with data on
charged hadron multiplicity spectrum for $|\eta|<0.5$, 1.0, 1.5, 2.0 and 2.4 at $\sqrt{s}=7$ TeV \cite{CMS}.}
\end{center}
\end{figure}

\begin{figure}
\label{difdad010}
\vspace{2.0cm}
\begin{center}
%\vspace{-0.6cm}
\includegraphics[height=.60\textheight]{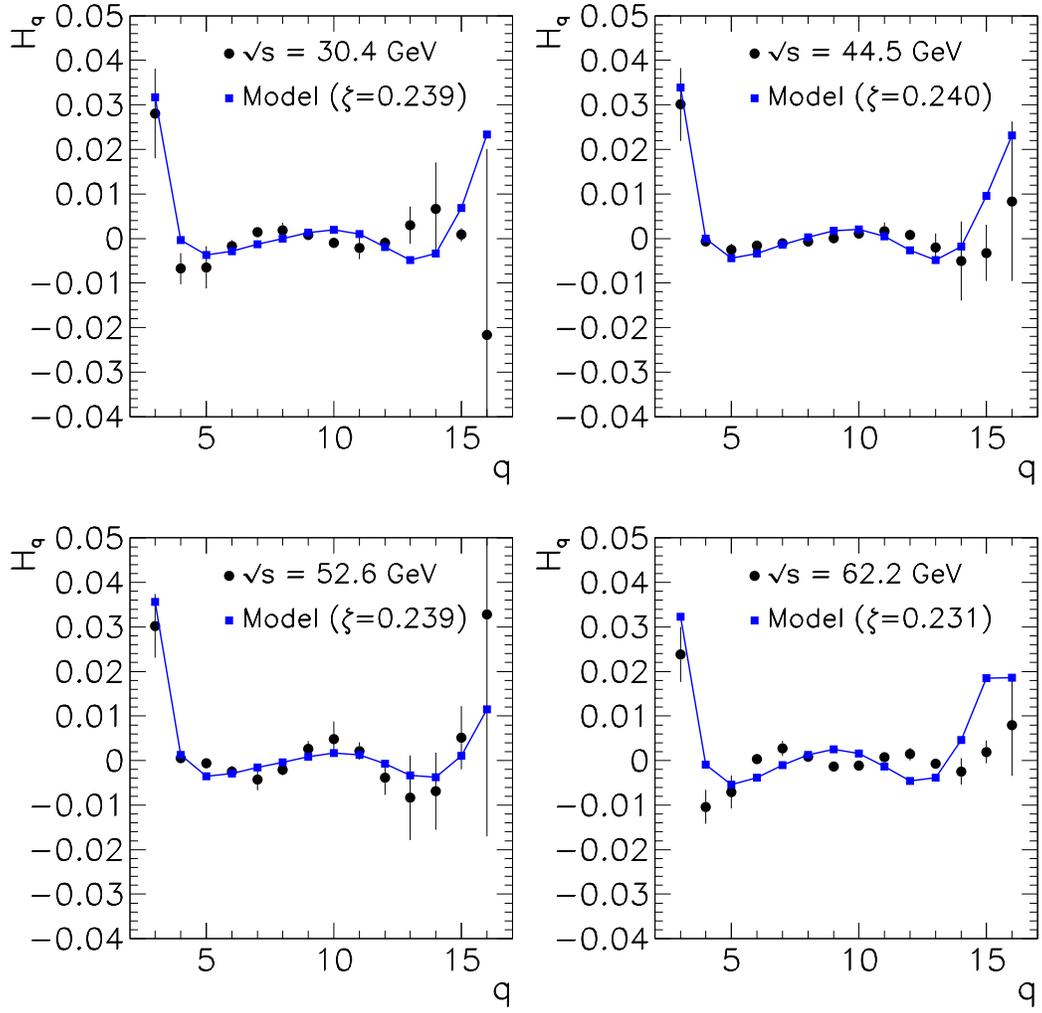}
\caption{Theoretical and experimental plots of $H_{q}$ against $q$ for $pp$ at ISR energies.}
\end{center}
\end{figure}

\begin{figure}
\label{difdad011}
\vspace{2.0cm}
\begin{center}
%\vspace{-0.6cm}
\includegraphics[height=.60\textheight]{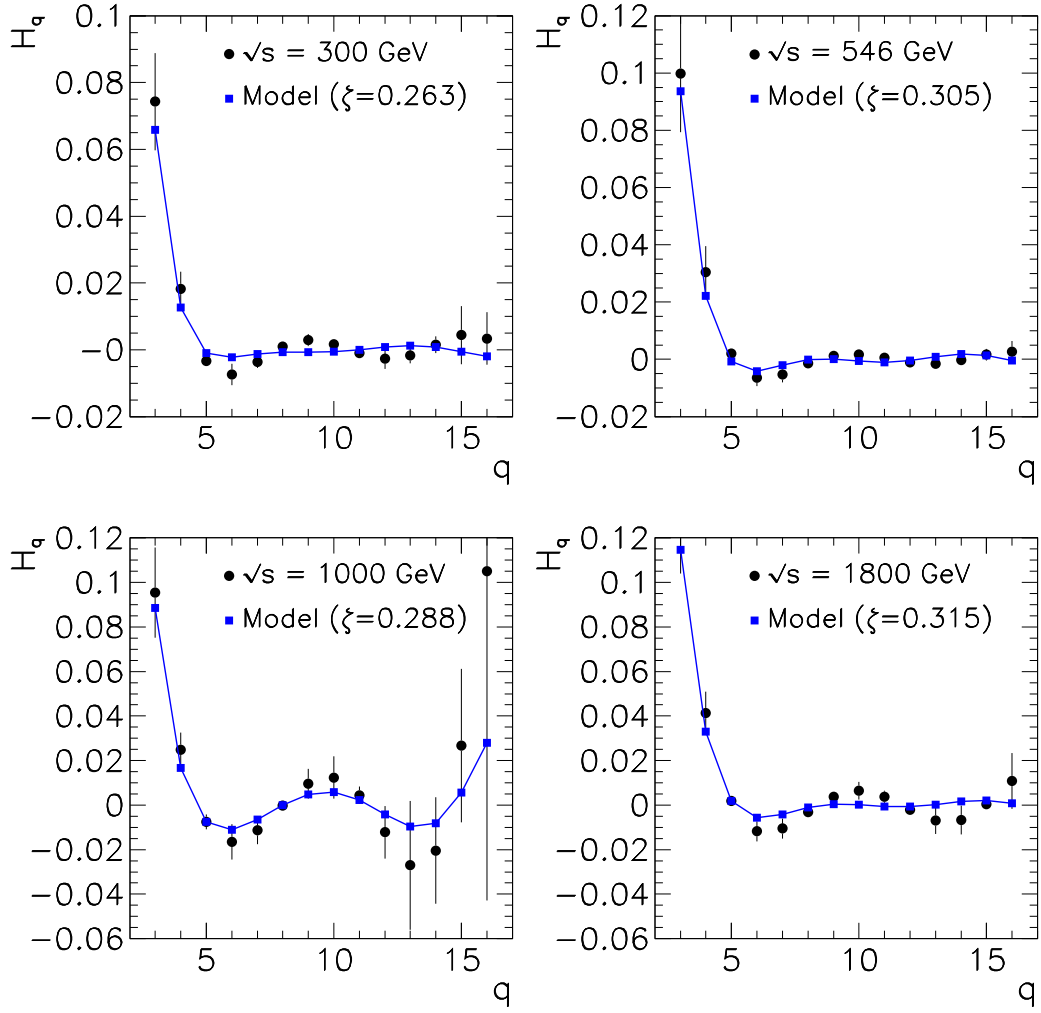}
\caption{Theoretical and experimental plots of $H_{q}$ against $q$ for $\bar{p}p$ at Collider energies. }
\end{center}
\end{figure}

\begin{figure}
\label{difdad012}
\vspace{2.0cm}
\begin{center}
%\vspace{-0.6cm}
\includegraphics[height=.60\textheight]{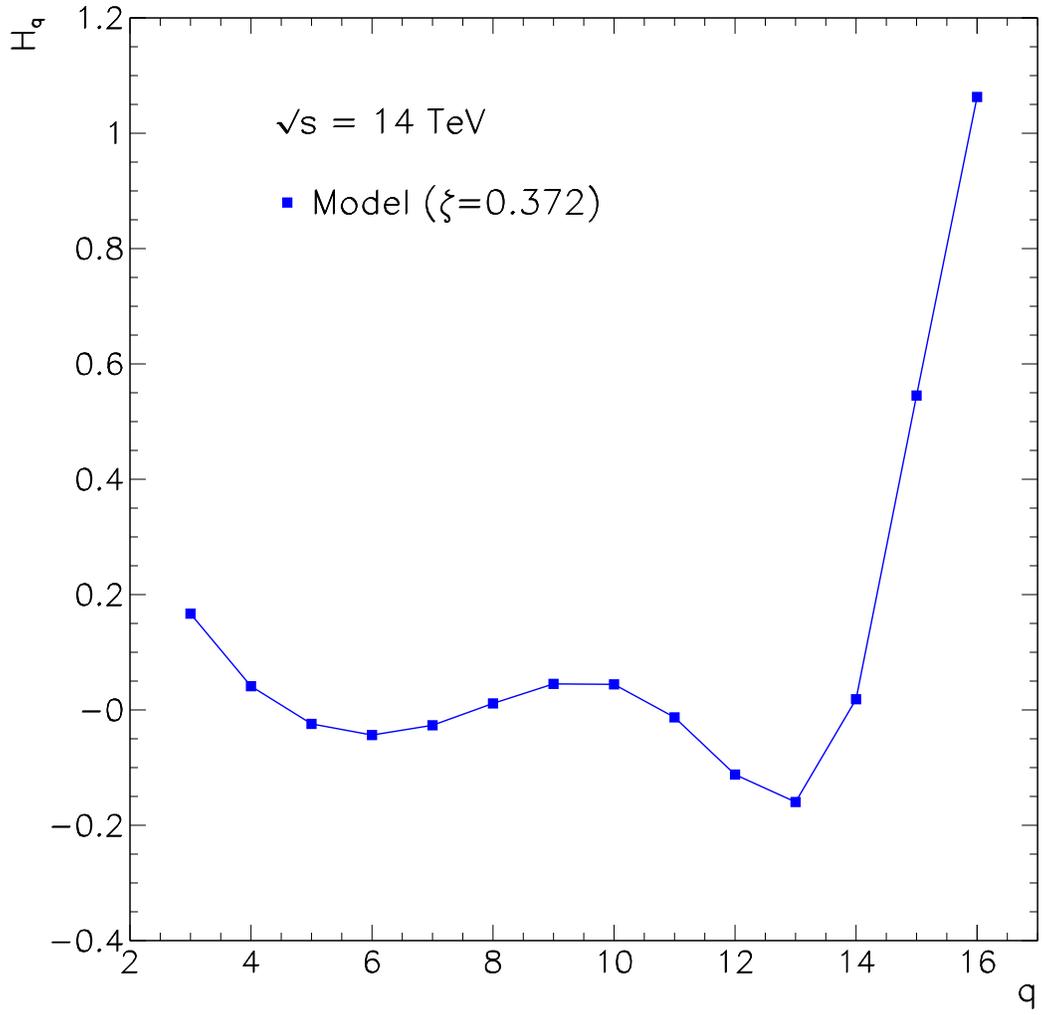}
\caption{Theoretical prediction for the ratio of the normalized factorial cumulant $K_{q}$ to the factorial moment $F_{q}$
at $\sqrt{s}=14$ TeV. }
\end{center}
\end{figure}

\end{document}